%% file: main.tex
\documentclass[twocolumn]{aastex6}

\pdfoutput=1

\usepackage{booktabs}
\usepackage{multirow}

\usepackage{amsmath}
\usepackage{amsfonts}
\usepackage{amssymb}
\usepackage{wasysym}
\usepackage{comment}
\usepackage{booktabs}
\usepackage{ulem}

\begin{document}

\title[GCM Rotation]{Climates of Warm Earth-like Planets II: Rotational `Goldilocks' zones for fractional habitability and silicate weathering}

\author{Tiffany Jansen\altaffilmark{1}, Caleb Scharf\altaffilmark{1}, Michael Way\altaffilmark{2}$^,$\altaffilmark{3}, Anthony Del Genio\altaffilmark{2}}

\shorttitle{Rotation, Weathering, and Habitability}
\shortauthors{Jansen et al.}

\affil{jansent@astro.columbia.edu}
\altaffiltext{1}{Department of Astronomy, Columbia University, 550 W 120th St., New York, NY 10027, USA}
\altaffiltext{2}{NASA Goddard Institute for Space Studies, 2880 Broadway, New York, NY 10025, USA}
\altaffiltext{3}{Department of Physics and Astronomy, Uppsala University, Uppsala, 75120, Sweden}

\begin{abstract}
    Planetary rotation rate has a significant effect on atmospheric circulation, where the strength of the Coriolis effect in part determines the efficiency of latitudinal heat transport, altering cloud distributions, surface temperatures, and precipitation patterns. In this study we use the ROCKE-3D dynamic-ocean general circulation model to study the effects of slow rotations and increased insolations on the `fractional habitability' and silicate weathering rate of an Earth-like world. Defining the fractional habitability $f_{h}$ to be the percentage of a planet's surface which falls in the $0\le T\le 100^{\circ}$C temperature regime, we find a moderate increase in $f_{h}$ with a 10\% and 20\% increase in insolation and a possible maximum in $f_{h}$ at sidereal day-lengths between 8 and 32 times that of the modern Earth. By tracking precipitation and run-off we further determine that there is a rotational regime centered on a 4-day period in which the silicate weathering rate is maximized and is particularly strongly peaked at higher overall insolations. Because of weathering's integral role in the long-term carbonate-silicate cycle, we suggest that climate stability may be strongly affected by the anticipated rotational evolution of temperate terrestrial-type worlds, and should be considered a major factor in their study. In light of our results we argue that planetary rotation period is an important factor to consider when determining the habitability of terrestrial worlds.

\end{abstract}


\section{Introduction}
\label{sec:intro}
\input{introduction.tex}

\section{Methods}
\label{sec:methods}
\input{methods.tex}

\section{Results}
\label{sec:results}
\input{results.tex}

\section{Discussion}
\label{sec:discussion}
\input{discussion.tex}

\section{Conclusion}
\label{sec:conclusion}
\input{conclusion.tex}

\acknowledgments
This research was supported by the NASA Astrobiology Program through participation in the Nexus for Exoplanet System Science and NASA Grant NNX15AK95G. This work was also supported by the NASA Goddard
Space Flight Center Sellers Exoplanet Environments Collaboration.

We would also like to thank Igor Aleinov and Nancy Kiang for their helpful discussions on the runoff and soil related diagnostics of the ROCKE-3D GCM.

\bibliography{mainbib}

\end{document}

%% file: introduction.tex
Modeling the atmospheres and surface environments of Earth-like worlds is a highly complex problem with potentially rich rewards in the field of exoplanetary science. Deepening our understanding and physical intuition of the forces governing a planet's climate not only allows us to speculate about the diversity of conditions and potential sustainability of life beyond our solar system, but also grants insight into the past, present, and future Earth.

The most complete method of climate modeling is the application of three-dimensional General Circulation Models (GCMs). GCMs specialize in the diagnostics of atmospheric dynamics, with the ability to track wind velocities, changes in latent heat, differential albedos, and in the case of the GCM used in this study (see below and \citealt{way:2017}), changes in oceanic energy transport and ground hydrology. Each GCM is unique in its capabilities  and computational efficiency depending on the  desired complexity of the model (see \citealt{eyring:2016} and references therein for information on recent climate model comparison efforts). 

\begin{figure}
    \centering
    \includegraphics[width=\linewidth]{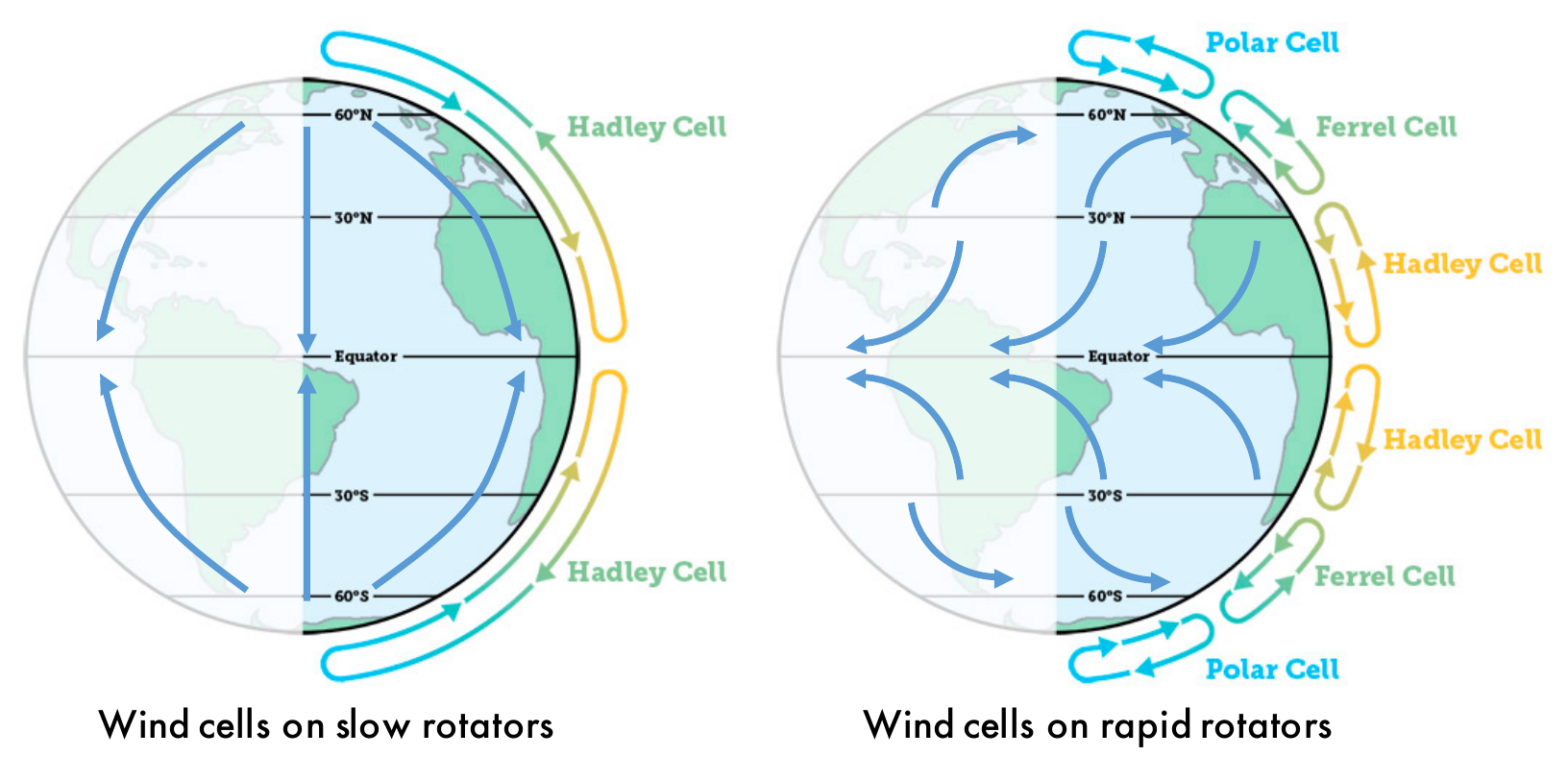}
    \caption{An illustration of the influence of the Coriolis effect on atmospheric circulation (arrows on the globe) for slowly rotating planets and more rapidly rotating planets. For reference, the wind cells on Earth are more similar to the diagram on the right.}
    \label{fig:coriolis}
\end{figure}

The rate at which a planet rotates about its axis has a significant impact on its atmospheric circulation. At the most fundamental level, the number and latitudinal extent of wind cells on a planet is determined in part by the strength of the Coriolis effect (Figure \ref{fig:coriolis}), which itself is determined by the planet's rotation period. \cite{delgenio:1987} used the Goddard Institute for Space Studies Model 1 GCM to show that for terrestrial bodies rotating more slowly than the Earth, the Hadley cell extends toward the poles and becomes the primary source of large-scale heat transport. This results in a more uniform temperature distribution across latitudes. The latitudinal extent of the Hadley cell also plays a significant role in large scale precipitation patterns across a surface, and is (for example) responsible for arid desert regions occurring in northern Africa rather than in continental Europe here on Earth.

However, determining the precise origins and distributions of rocky planet rotations is a fundamental and challenging question in planetary and astronomical science. Because we can only observe the rotation rates of rocky planets in our own solar system, for now we must rely on theoretical predictions to estimate the ranges of possible rotation rates. For example, \cite{miguel:2010} suggest the initial rotation periods of rocky planets range from 10 to 10,000 hrs. Other work suggests the upper limit on a planet's initial rotational velocity is bound by the magnitude of its surface gravity, i.e. the critical angular velocity of rotational stability \citep{kokubo:2007}. 

The rotation period of a planet at a given time is also dependent on its evolutionary path and frequency of encounters with exchanges in angular momentum, including star-planet and planet-moon tidal dissipation. For example, \cite{barnes:2016} shows that for a Moon-less Earth with an initial rotation period of 3 days, stellar tides would cause its rotation to become synchronous within 4.5 Gyr. Evolution toward synchronized rotation due to tides is even swifter for potentially temperate planets around lower mass stars, with the spin-down rate following $d\Omega/dt \propto M_*^{2}/a^6$, where $a$ is the orbital semi-major axis \citep{goldreich:1966}. For an Earth-like planet in the habitable zone of a Gliese 581-like M dwarf star, \cite{heller:2011} show that the evolution time scale for equilibrium rotation is about 100 Myr. In light of these studies, the shape of the probability distribution of rocky planet rotation rates is debatable, although its range is most certainly wide. We do however expect a trend towards overall slower rotation rates with age due to star-planet tidal interactions.

 \cite{yang:2014} used the Community
Atmosphere Model version 3.1 GCM \citep{collins:2004} to show that clouds tend to congregate around the substellar point of slowly rotating planets, thus increasing the planetary albedo and decreasing the surface temperature at that point, subsequently extending the inner habitable zone to smaller orbits. However, it is important to note that most of the \cite{yang:2014} study was conducted on an aquaplanet configuration, and because of the different heat capacities and albedos of land and ocean, we expect a significant land mass to have a non-zero effect on the overall climate. 

In our study we adopt the continental configuration of the Earth to study the effects of slow rotations and increased insolations on the climate of an Earth-like world using the Resolving Orbital and Climate Keys of Earth and Extraterrestrial Environments with Dynamics (ROCKE-3D) GCM developed in multi-institutional collaboration at the Goddard Institute for Space Studies (GISS) \citep{way:2017}. This model is further differentiated from those used in previous rotation studies in that ROCKE-3D uses a fully coupled dynamic ocean scheme rather than the static ocean model often adopted.

We investigate both the fractional habitability and silicate rock weathering of an Earth-like terrestrial world as key probes of short- and long-term climate states and stability. It is uncontroversial to state that the now-classical `zeroth-order' marker of habitability given by a global mean surface temperature \citep{kasting:1993, kopparapu:2013} is only a broad indicator of terrestrial planet environment; useful for a first pass evaluation of a system, but too basic for either predicting observable characteristics or potential for life. However, extending the parameterization of `habitability' to higher orders (e.g. details of atmospheric chemistry, hydrological features, orbit and rotation, planet composition and geophysics, spatially-resolved climate) quickly becomes cumbersome and uninformative in the absence of observable constraints on many variables. A variety of first-order or even second-order markers have been used to further refine habitability predictions (e.g. \citealt{spiegel:2008,schulze:2011}). In this study we perform a first-order analysis similar to that of \cite{spiegel:2008} to evaluate the `habitability' of rocky worlds with rotation periods slower than that of modern Earth.

Furthermore, although the temperature range of 0-100 $^{\circ}$C is not a strict set of boundaries for biochemistry and biological function (\citealt{rothschild:2001} and references therein) it does nonetheless encompass the temperature range over which known biochemistry takes place at high efficiency for modern Earth atmospheric surface pressures \citep{gillooly:2001, rothschild:2007}. We also emphasize that because this temperature range is directly related to H$_{2}$O phase changes on a planet with an Earth-like atmospheric composition and surface gravity, it is centrally important to changes in surface albedo. These phase changes are also critical to the global climate state due to the contribution of latent heat to the overall energy transport in the atmosphere. In other words, in this present work we treat the term `habitability' as a convenient label, primarily for evaluating the physical and chemical environment of a rocky planet.

In Section \ref{sec:methods} we describe the models used in this study, and define the methods we use to determine the fractional habitability and silicate weathering rate from the model output. In Section \ref{sec:results} we provide results on the behavior of surface temperature, fractional habitability, precipitation, and the silicate weathering rate for longer rotation periods. In Section \ref{sec:discussion} we discuss the implications of these results, and present a brief conclusion in Section \ref{sec:conclusion}.

%% file: methods.tex
\subsection{GCM model}

The ROCKE-3D general circulation model was developed from the NASA Goddard Institute for Space Studies' preexisting ModelE2 GCM \citep{schmidt:2014}, whose publicly available model results for Earth climate studies continue to be utilized by the international community. ROCKE-3D was created for the purpose of extending the capabilities of ModelE2 toward modeling exoplanet atmospheres and rocky solar system planet atmospheres. For this study we use ROCKE-3D's dynamic ocean capabilities, which are fully coupled to the model atmosphere. A brief description of the key elements of the models used in this study is below. For a more comprehensive description of ROCKE-3D, we refer the reader to \cite{way:2017}. 

Each model planet in this study is one Earth radius in size and has a topography and continental/oceanic distribution similar to that of the Earth. For dynamic simplicity, we set the eccentricity and obliquity of our model planets to zero. The surface resolution is $4^{\circ}$ by $5^{\circ}$ in latitude and longitude, respectively. For the fully coupled dynamic ocean we use oceanic layers of varying thickness, with an overall depth of 1129.3 meters. The atmosphere spans nearly one bar of pressure from 984 to 0.139 mb over 40 vertical layers of varying thickness.

The dominant atmospheric constituent in each model is N$_2$, where CO$_2$ and CH$_{4}$ are included at 400 and 1.00 parts per million, respectively. H2O is given a modern-Earth profile at model start, which changes depending on the atmospheric circulation, evaporation, and precipitation in each simulation. It should be noted that all runs have oxygen-free atmospheres, and no ecosystem models are implemented, i.e. vegetation and other biological sources are not included in these simulations.

ROCKE-3D uses the Suite of Community
Radiative Transfer (SOCRATES) radiation scheme to solve the two-stream approximated radiative transfer equation \citep{edwards:1996a, edwards:1996b}. In this study we use the present-day Solar spectrum to weight the short-wave component of the planetary flux, although SOCRATES does have the capability to use a variety of stellar spectra. Land albedo is set to 0.2 at model start, although this can be changed by snow accumulation and soil wetness on the surface.

To study the effects of rotation rate on the surface temperature of an Earth-like world, we analyze the results of nine simulation runs with rotations slower than (and equal to) that of present-day Earth. These rotations are $t_{rot}=$ 1, 2, 4, 8, 16, 32, 64, 128, and 256 times the sidereal day length of present-day Earth. In solar days, these correspond to 1, 2, 4.03, 8.16, 16.7, 35.0, 76.6, 191, and 848 times the solar day length of present-day Earth. Additionally, we consider the effects of increasing the insolation S0 by 10\% and 20\% for each rotation period, where S0 = 1365.3 W m$^{-2}$, the insolation of modern-day Earth. Specific insolations will be referenced using the nomenclature 00t$_{rot}$X1.nS0, where 1.n indicates the insolation multiplier for n = 0, 1, or 2. For example, the simulation with a rotation period of 32 days and an insolation of 1.2S0 will be referenced as 032X1.2S0.

The GCM data presented in this study have been averaged over ten model years post hydrological and radiative equilibrium, which are reached within an order of $10^{2}$ simulated years. For a more in depth description of the simulations in this study, see \cite{way:2018b}.

\subsection{Fractional Habitability}

Here we define the `fractional habitability' $f_h(t_{rot})$ as the fractional surface area which maintains a mean temperature between 0 and 100 $^{\circ}$C at 984 mb of atmospheric pressure, averaged over at least one orbital period and diurnal cycle for a specific orbital period $t_{rot}$,
\begin{equation}
    f_{h}(t_{rot}) = \frac{1}{A_{tot}}\sum_{i=1}^{n_{\theta}}\sum_{j=1}^{n_{\phi}}H(\theta_{i}, \phi_{j}, t_{rot})A_{i,j}
\end{equation}
for total surface area $A_{tot}$, specific latitude grid $\theta_{i}$ and longitude grid $\phi_{j}$, number of latitude and longitude grids $n_{\theta}$ and $n_{\phi}$, respectively, and the area of the $i,j$th grid cell $A_{i,j}$, where we have adopted the `habitability function' $H(\theta, \phi, t_{rot})$ from \cite{spiegel:2008}, defined to be
\begin{equation}
   H(\theta, \phi, t_{rot})=
   \begin{cases} 
      1 & \text{if } 0 \leq T(\theta, \phi, t_{rot})[^{\circ}\text{C}] \leq 100 \\
      0 & \text{else}
   \end{cases}.
\end{equation}

We suggest that this is a useful 1st-order probe of climate state (habitability, see \S1) for our simulations, for which seasonal variation is minimal. In a more general case where seasonal variation does occur, we would be motivated to consider both the surface area that is always capable of supporting liquid water, and the surface area that can support liquid water for part of a single orbit, i.e. the seasonal fractional habitability, as is done in \cite{spiegel:2008}. Because our simulations are of planets with zero obliquity and zero orbital eccentricity, season variations are not present.

Instead, $f_h(t_{rot})$ captures key aspects of simulation-to-simulation variations in cloud cover and albedo across the surface, as well as variations in heat transport. In other words, $f_h(t_{rot})$ contains more information about the state of a climate than the globally averaged mean temperature, yet is sufficiently simple to allow us to build a meaningful physical intuition for the conditions of systems that have many sensitivities and underlying parameter dependencies.

\subsection{The Silicate Weathering Rate}

The carbonate-silicate cycle plays an important role in the stabilization of the climate of water-bearing terrestrial worlds. An element of this cycle is the silicate weathering process (e.g. \citealt{kasting:1993}), in which CO$_{2}$ is drawn from the atmosphere as carbonic acid in precipitation. When this acid contacts silicate rock, the silicates are chemically dissolved (with reaction rates dependent on temperature and acid concentrations) and the products are carried into the oceans via runoff. Eventually the products of silicate weathering, such as calcite (CaCO$_3$), sink to the ocean floor and become subducted into the mantle, where carbonate metamorphism eventually transforms them back into CO$_{2}$ which may re-enter the atmosphere diffusely or via volcanic outgassing. This process is commonly associated with the regulation of Earth's climate close to a state where surface liquid water is abundant \citep{berner:1983, walker:1981}. For example, a decrease in solar forcing would result in a drop in surface temperature and precipitation, leading to a build up of atmospheric CO$_{2}$ and therefore an increased greenhouse effect, effectively re-warming the surface. If surface temperatures and precipitation patterns are significantly altered by planetary rotation rate, it is critical to our understanding of long-term exoplanetary climate stability to contemplate how the silicate weathering rate is impacted by rotation. 

We adapt the method of \cite{walker:1981} to model the total silicate weathering rate $W$ from the ROCKE-3D output, normalized by the total weathering rate $W_{0}$ of the 001X1.0S0 model, i.e. the 1-solar insolation, 1-day rotation period model,
\begin{equation}\label{eq:weathering}
    W/W_{0} = \frac{\sum_{i,j}R_{i,j}\ \text{exp}\left(\frac{T_{i,j}}{17.7}\right)}{\sum_{i,j}R_{0,i,j}\ \text{exp}\left(\frac{T_{0,i,j}}{17.7}\right)}
\end{equation}
where $R_{i,j}$ is the runoff of precipitation over bare soil on the surface and over soil underground in the $i,j$th cell, weighted by the fraction of soil in that cell. $T_{i,j}$ is the soil temperature in the $i,j$th cell (in Kelvin) averaged across layers of soil with a total depth of 3.5 meters, and weighted by the thickness of each soil layer and the area of the grid cell. The subscript ``$0$'' indicates the values from the 001X1.0S0 model.

Our representation of the relative weathering rate deviates from Eq. 1 of \cite{walker:1981} in a number of important ways. First, we exclude the expression for the dependence on the partial pressure of CO$_{2}$ (i.e. $\left(P/P_{0}\right)^{0.3}$) because the abundances of CO$_{2}$ are fixed across all models of varying rotations and insolations, and we assume the partial pressures of CO$_{2}$ remain constant over the relatively short timescales of the simulation runs. Thus $\left(P/P_{0}\right)^{0.3}$ is simply equal to unity for all cases explored in this study. 

The temperature scaling factor of 17.7 in the exponent of Equation \ref{eq:weathering}\ is chosen according to the following rationale. \cite{walker:1981} used experimental results of the temperature dependence of the silicate weathering rate \citep{lagache:1965,lagache:1976} together with conservative results from climatological models to get a scaling factor of 13.7 as an estimate of the dependence of runoff on temperature \citep{wetherald:1975, menabe:1980a, menabe:1980b}. This is the value used in many silicate weathering studies (e.g. \citealt{sleep:2001, rushby:2018}), however, because ROCKE-3D gives runoff as a diagnostic, our study does not require an estimate of the dependence on temperature as a proxy for runoff and we therefore retain the temperature scaling factor of 17.7 determined solely by the experimental results of \cite{lagache:1965,lagache:1976}.  

%% file: results.tex
\subsection{Fractional Habitability}

\begin{figure}
    \centering
    \includegraphics[width=\linewidth]{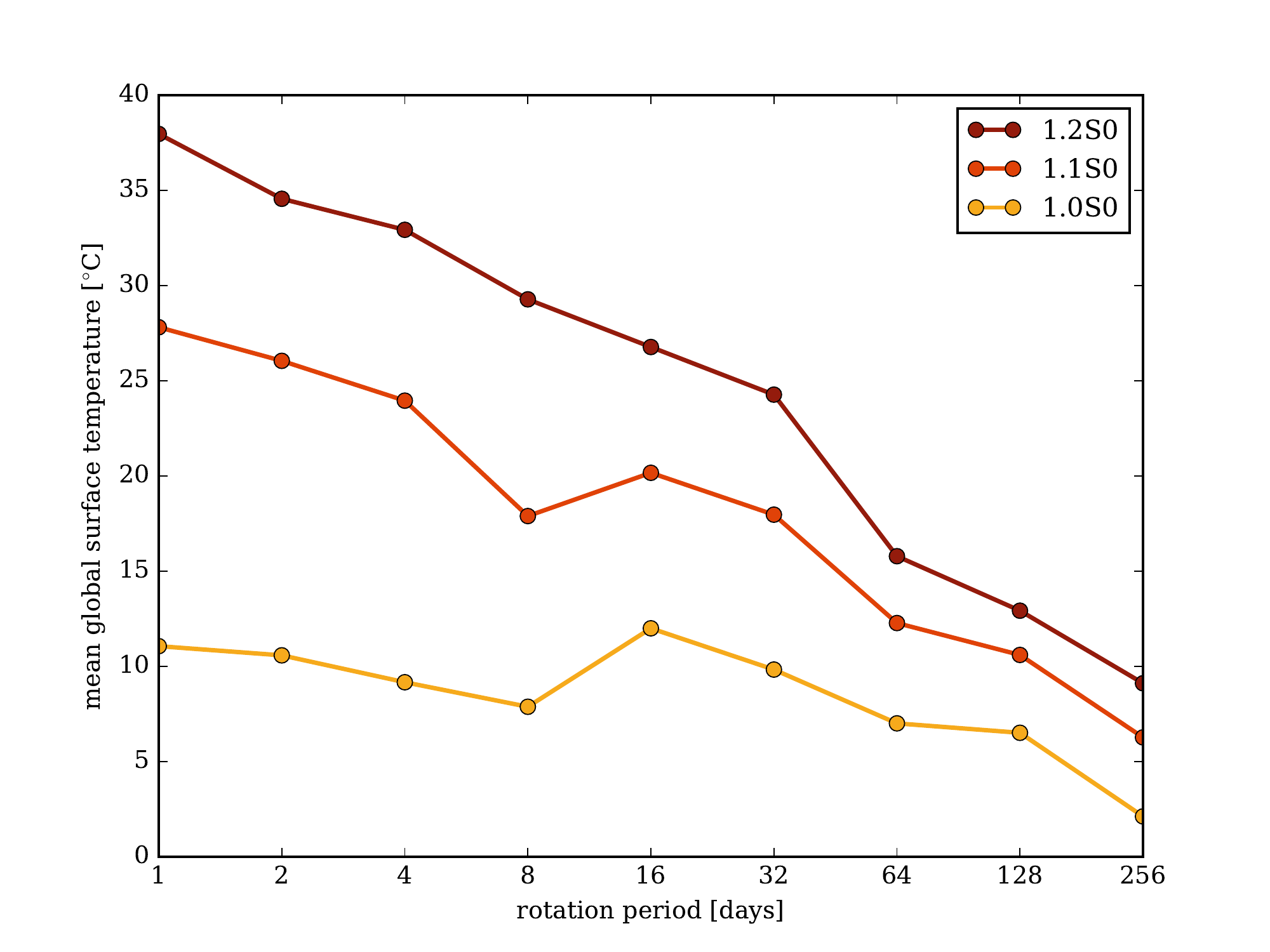}
    \caption{Mean global surface temperature in $^{\circ}$C for each model as a function of rotation period (in sidereal days), with insolations (in solar insolations S0) represented by lines of different color.}
    \label{fig:avg_glob_tsurf}
\end{figure}

As shown by \cite{way:2018b}, the ROCKE-3D runs provide evidence that the average global surface temperature tends to decrease with increasingly long days (i.e. longer rotation periods) for runs with 10\% and 20\% increases in insolation (Figure \ref{fig:avg_glob_tsurf}). This trend is clearest in the 1.2S0 runs due to the increased insolation initiating a global climate state that contains fewer regions capable of switching between ice-covered and ice-free for all rotations. The relationship between average global surface temperature and rotation period is less consistent for the solar insolation models, although the 064X1.0S0, 128X1.0S0, and 256X1.0S0 rotation models remain at lower average surface temperatures than the more rapidly rotating model planets. 

\begin{figure}
    \centering
    \includegraphics[width=0.5\textwidth]{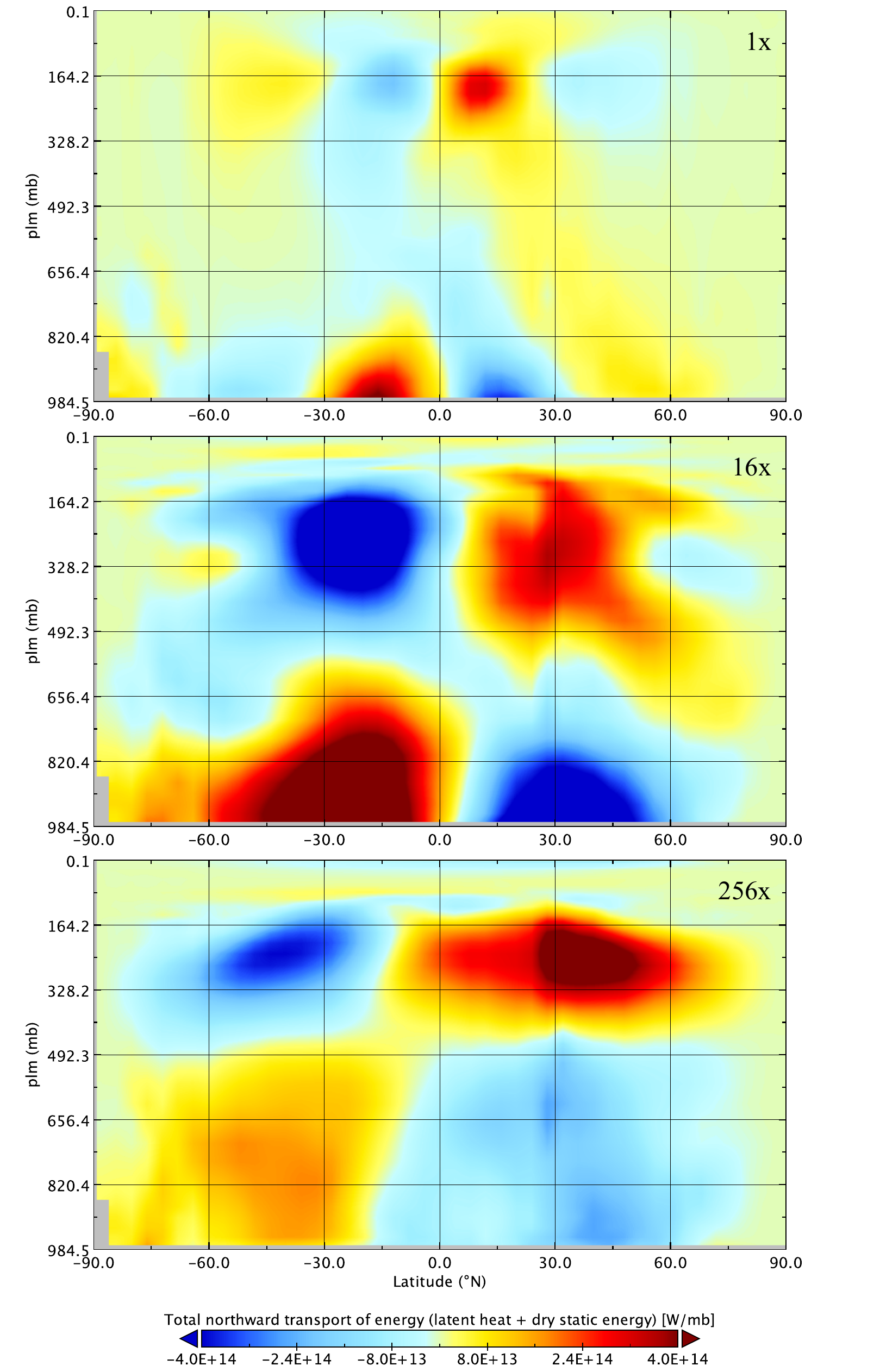}
    \caption{Total northward transport of energy as a sum of the dry static energy transport and latent heat transport in the solar insolation case for a different rotation period in each subfigure. From top to bottom these rotation periods are 1x, 16x, and 256x the sidereal day length of present Earth.}
    \label{fig:tot_nt_e}
\end{figure}

\begin{figure*}
    \centering
    \includegraphics[width=\textwidth]{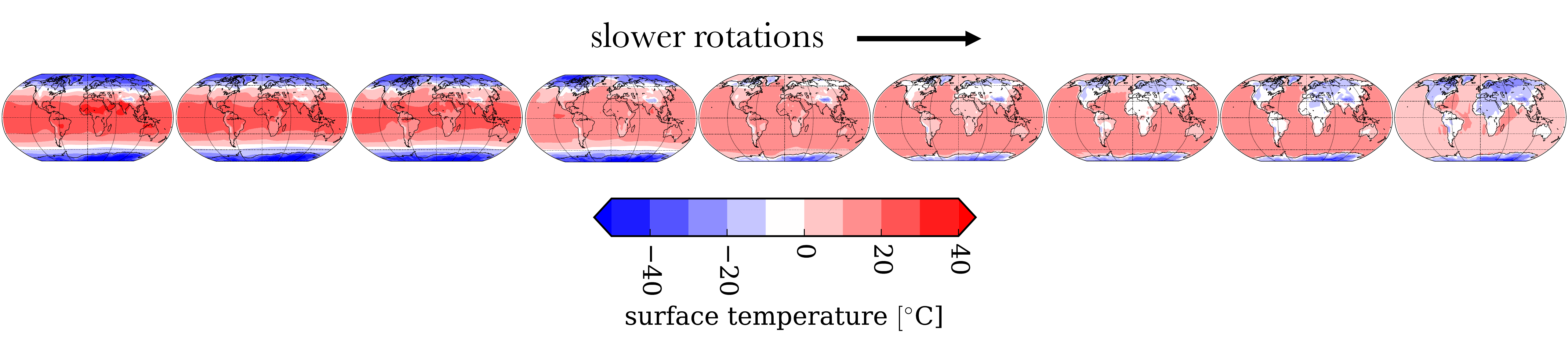}
    \caption{Surface temperatures for the 1.0S0 case (i.e. solar insolation) as a function of rotation shown on a Robinson projection of the globe. Rotation periods from left to right are 1x, 2x, 4x, 8x, 16x, 32x, 64x, 128x, 256x the sidereal day length of present Earth. Approximate continental configuration can be seen outlined in black. All red-colored regions are above freezing. }
    \label{fig:temps_globe}
\end{figure*}

Figure \ref{fig:tot_nt_e} shows the total northward transport of dry static energy and latent heat for a sample of rotation periods at 1.0 solar insolation. While the transport changes direction at lower altitudes as expected for the Hadley cell, the net energy transport is indeed poleward. This illustrates that heat is more efficiently transported toward the poles for slower rotating planets, causing the more uniform distribution of surface temperature across latitudes seen in Figure \ref{fig:temps_globe}. Due to the high thermal inertia of the model ocean which reaches depths of nearly 1300 meters, the oceans stay above their freezing point despite the tendency for land masses to slip below 0$^{\circ}$C, especially for model planets with the longest day lengths where land masses cool significantly on the night side.

\begin{figure}
    \centering
    \includegraphics[width=\linewidth]{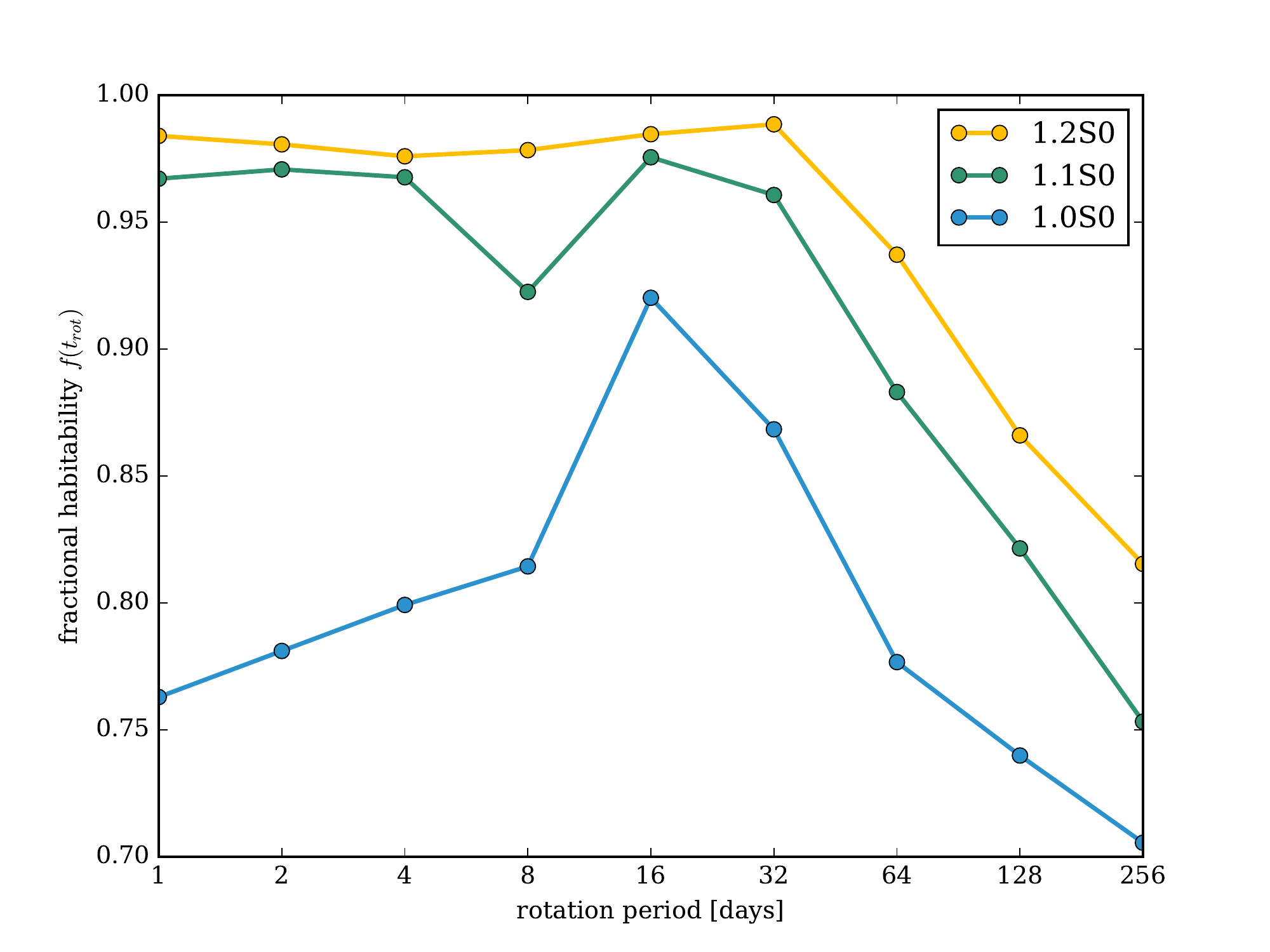}
    \caption{Fraction of the total surface area where the surface temperature $T$ [$^{\circ}$C] lies in the range $0\le T \le 100$ as a function of rotation period (longer days to the right) and insolation S0.}
    \label{fig:hab_frac}
\end{figure}

The fraction of the surface which lies in the temperature range $0\leq T \leq 100 ^{\circ}$C for each rotation period and insolation can be seen in Figure \ref{fig:hab_frac}, and specific values of $f_{h}$ for each rotation and insolation are shown in Table \ref{table:hab_fracs}. We find moderate increases in fractional habitability $f_{h}$ across all rotations for 10 and 20\% increases in insolation, which follows intuitively from the increase in mean global surface temperature with insolation (Figure \ref{fig:avg_glob_tsurf}).

For the 1.0S0 insolation case, there is an intriguing rise and decline in fractional habitability as a function of rotation, with a defined peak of $f_{h}(16)=0.92$ at the 16-day rotation period model. This is an additional 12\% of temperate surface area compared to the average value of $\left<f_h\right>=0.80$ for 1.0S0 insolation. This peak in fractional habitability at intermediate rotation periods is due to the interplay between two dynamical regime changes: one between a quasi-geostrophic state and a quasi-barotropic state, and another in which diurnal heating contrasts begin to dominate.

In the transition from a quasi-geostrophic state to a quasi-barotropic state, the Hadley cell extends to higher latitudes (Figure \ref{fig:tot_nt_e}) and the poleward temperature gradient decreases (see Figure 9 in \citealt{way:2018b}). This poleward extension of the relatively warm equatorial air effectively increases the surface area of `habitable' temperatures. This particular transition occurs when the Rossby radius of deformation approaches the size of the planet \citep{delgenio:1983, edson:2011, showman:2013}. For an Earth-sized planet, this occurs at a rotation period between 8 and 16 days  \citep{delgenio:1987}. 

The second dynamical transition occurs when the dominant circulation switches from an equator-to-pole motion to a day-to-night motion as the length of the solar day grows $\gtrsim$ the radiative relaxation time scale of the planet, which for our Earth-like model is on the order of 1-2 months \citep{way:2018b}. Therefore the length of the solar day becomes longer than our models' radiative relaxation time in the 32 and 64 day rotation period interval, at which point the diurnal temperature contrast begins to increase significantly. As the average nightside temperature decreases with increasing rotation periods, the total area in which water can remain in liquid form at the surface is effectively diminished.

The rotation period at which the peak in fractional habitability occurs is therefore a result of the overlap of these dynamical transitions in the rotation period domain --- in the 16-32 day interval, both the equator and poles are sufficiently warm while the nightside does not yet grow too frigid.

A common trend for every insolation is the steep abatement in fractional habitability for rotation periods greater than 16 and 32 days. In addition to the increase in diurnal temperature contrast, we speculate that this occurs after the efficiency of the poleward transport of heat from the equator has been maximized and an increase of cloud coverage at the substellar point works to reflect more incoming radiation as rotation period increases. Because we are unable to track the exact location of the substellar point in these particular simulation runs, our speculation cannot be addressed until we obtain averages from finer time sampling of the models in future studies.

\input{temp_hab_table.tex}

\subsection{Precipitation and Silicate Weathering Rate}

\begin{figure}
    \centering
    \includegraphics[width=\linewidth]{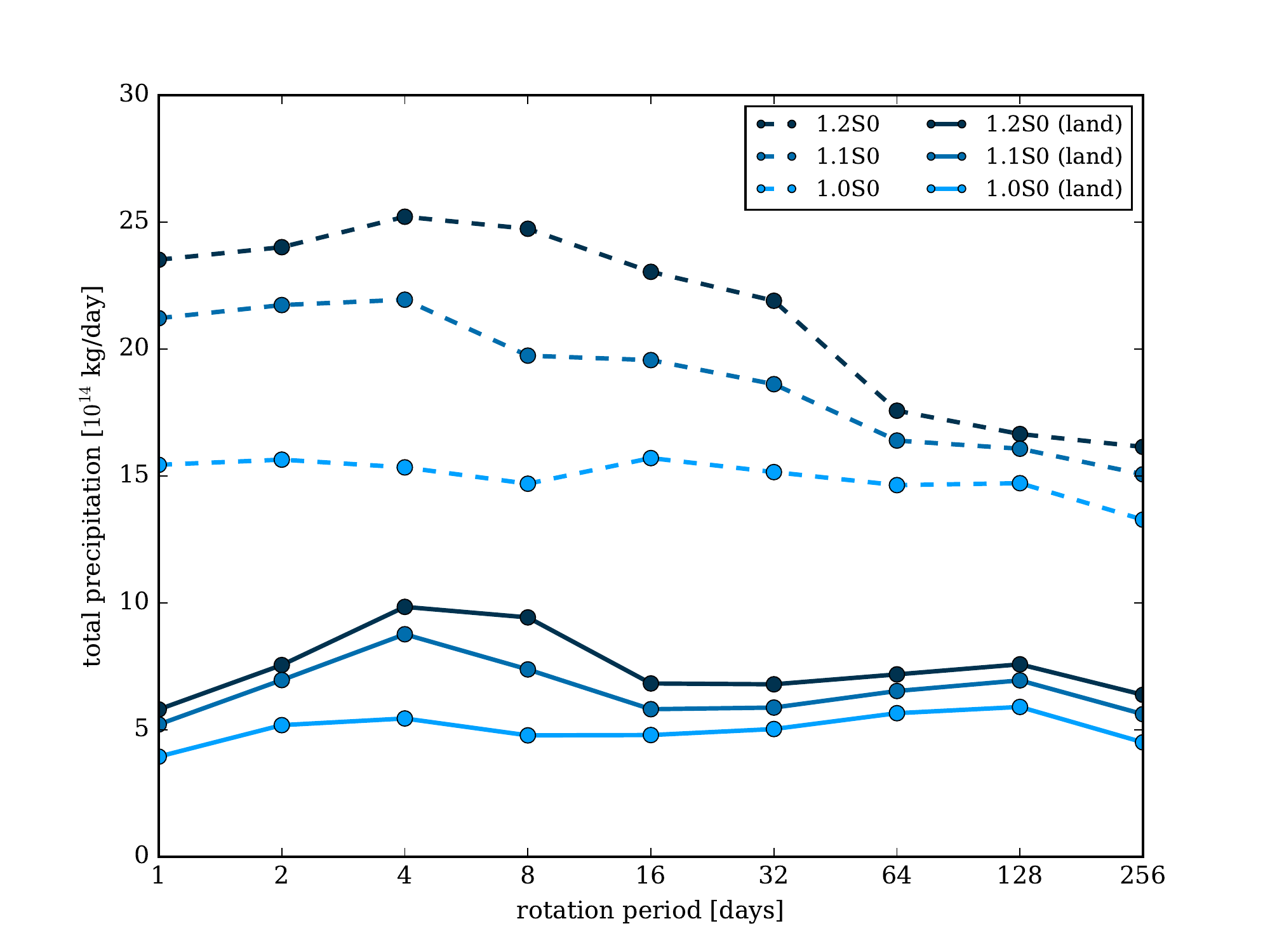}
    \caption{Total precipitation in kg day$^{-1}$ weighted by grid area as a function of rotation period and insolation. Solid lines indicate precipitation over bare soil (i.e. land) while dashed lines indicate precipitation over the entire surface. Models with different insolations are indicated by color.}
    \label{fig:total_precip}
\end{figure}

Surface temperatures rise with insolation and precipitation increases correspondingly. For a 10\% increase in insolation, the total precipitation summed over the planetary surface increases by an average of $(4\pm3)\times10^{14}$ kg/day across all rotation models, while for a 20\% increase in insolation, the total precipitation similarly increases by $(6\pm3)\times10^{14}$ kg/day compared to the 001X1.0S0 model. At these increased insolations, the relationship between rotation rate and precipitation is not as consistent as the relationship between rotation rate and surface temperature, despite their correlation. As can be seen in Figure \ref{fig:total_precip}, the four slowest rotating models (032X, 064X, 128X, 256X) have the lowest total precipitation, however, the 001X model experiences less total precipitation than the succeeding rotation models (002X, 004X) despite having a higher average global temperature. 

\begin{figure*}
    \centering
    \includegraphics[width=\textwidth]{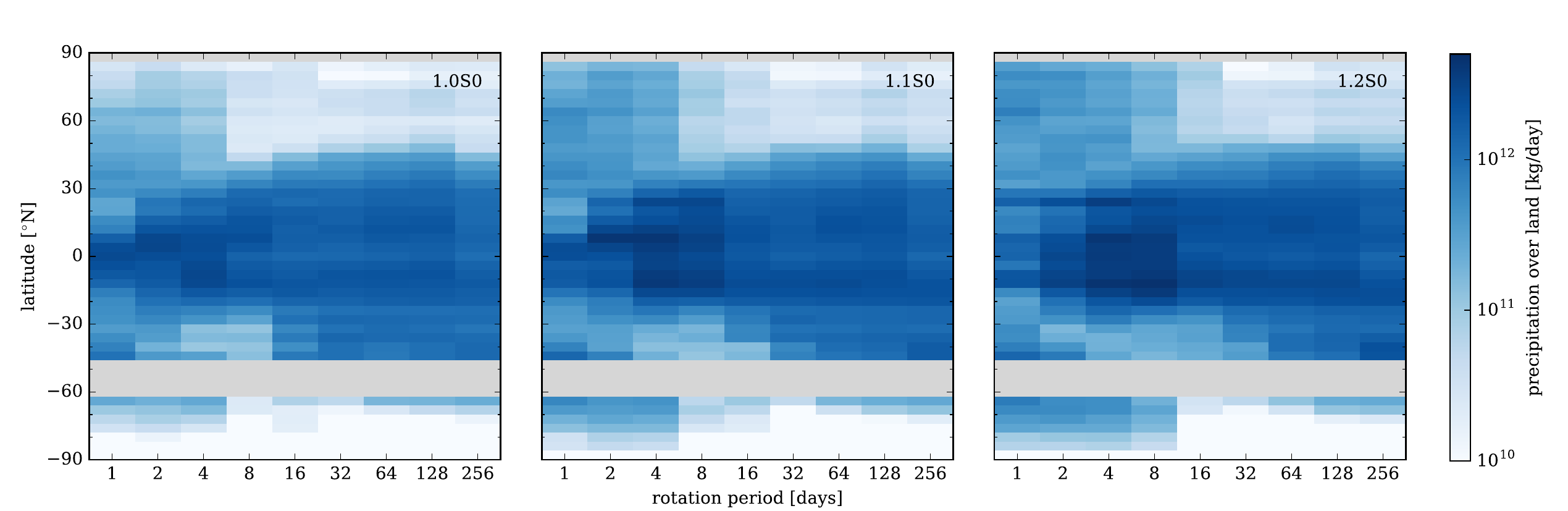}
    \caption{Kilograms per day of precipitation over land averaged across longitudes as a function of rotation and insolation S0. Rotation periods are in present Earth sidereal days. Grey regions mark latitudes which have no land masses.}
    \label{fig:precipitaion}
\end{figure*}

Another distinction of the more rapidly rotating models is the latitudinal extent of precipitation over land masses. At one solar insolation, the 001X, 002X, and 004X models show a noticeably greater amount of precipitation at higher latitudes than the models with increasingly long days (Figure \ref{fig:precipitaion}). For rotation periods $>$8 days, the bulk of the precipitation over land mass is concentrated within $50^{\circ}$ in latitude about the equator.

\begin{figure*}[p]
    \centering
    \includegraphics[width=0.8\linewidth]{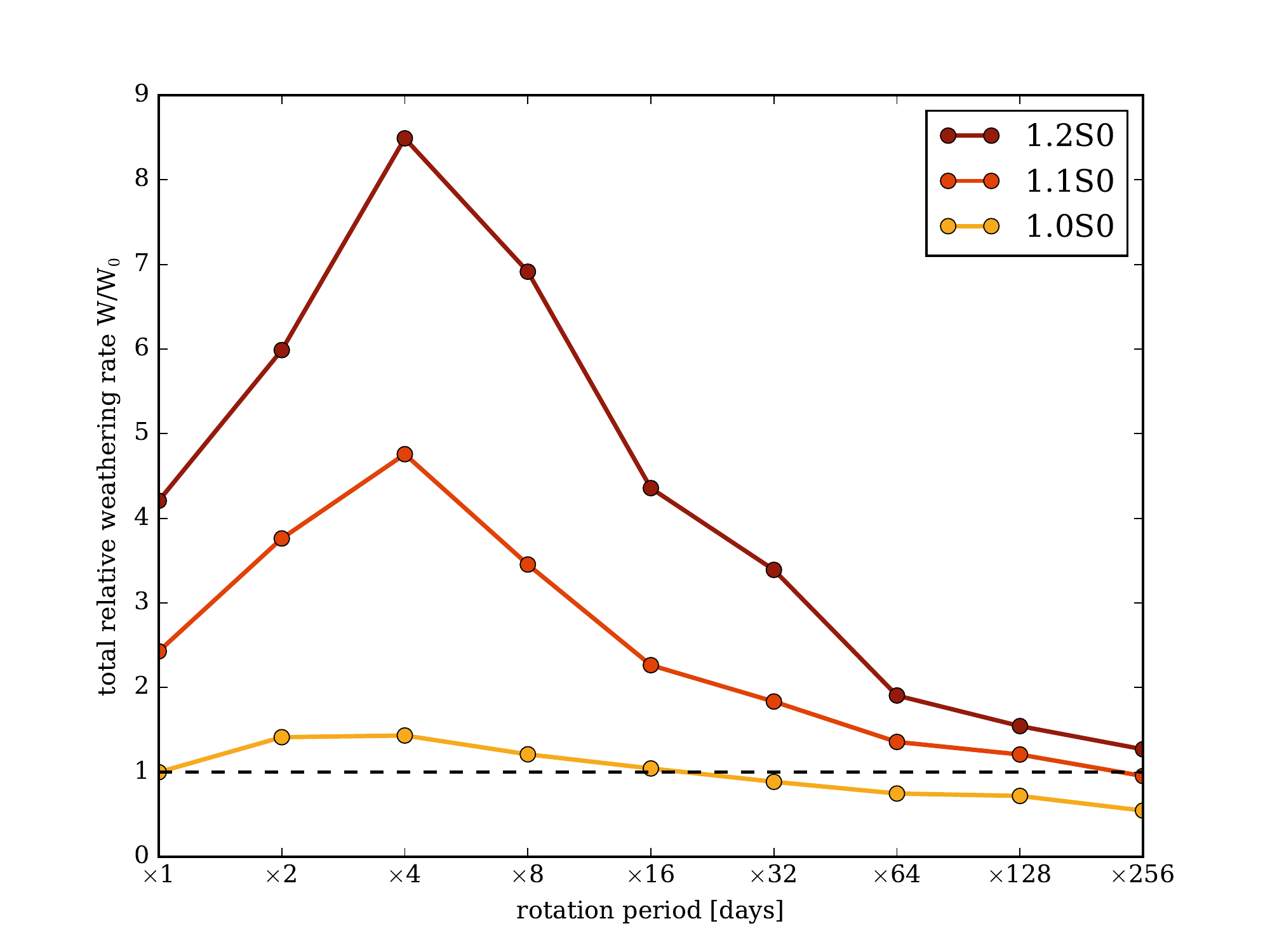}
    \caption{Total silicate weathering rate as a function of rotation and insolation relative to the weathering rate of the 1-day rotation 1.0 solar insolation case. The horizontal dashed line shows where the total weathering ratio between models is equal to 1.}
    \label{fig:weathering}
\end{figure*}

Increased precipitation over land will correlate with an increase in underground runoff (and surface runoff, depending on the topography of the location of precipitation). For a given temperature, an increase in runoff directly correlates to an increase in the rate of silicate weathering (Eq. \ref{eq:weathering}). This is reflected in Figure \ref{fig:weathering}, where the silicate weathering rate for a 10\% and 20\% increase in insolation increases by up to 8.5 times the rate of weathering of the 001X1.0S0 model.

As can also be seen in Figure \ref{fig:weathering}, there is a distinct growth and decline around a well defined peak in the relative weathering rates across all insolations, which reaches a maximum around a rotation period of 4 days. 
Despite having more precipitation over land (Figure \ref{fig:total_precip}), the models with rotation periods greater than 32 days experience decreased weathering rates relative to the 001X1.0S0 model. This is due to a decrease in soil temperature with increasingly long days.

\begin{figure*}[p]
\centering
    \begin{minipage}{0.48\linewidth}
    \centering
    \includegraphics[width=\linewidth]{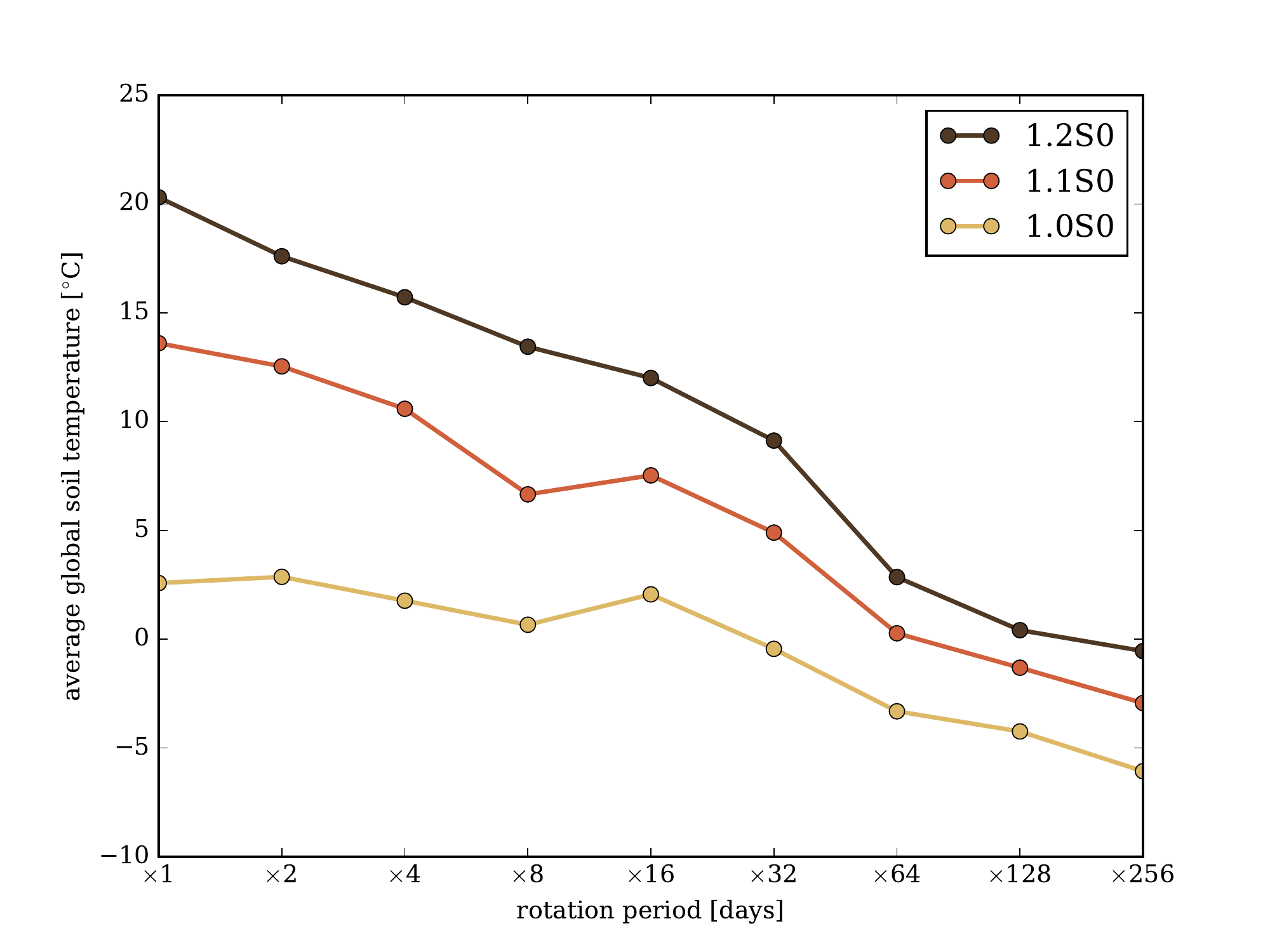}
    \caption{Average global soil temperature in Celsius weighted by soil thickness as a function of rotation period in days and insolation S0.}
    \label{fig:soil_temps}
    \end{minipage}
      \hspace{0.1cm}
    \begin{minipage}{0.48\linewidth}
    \centering
    \includegraphics[width=\linewidth]{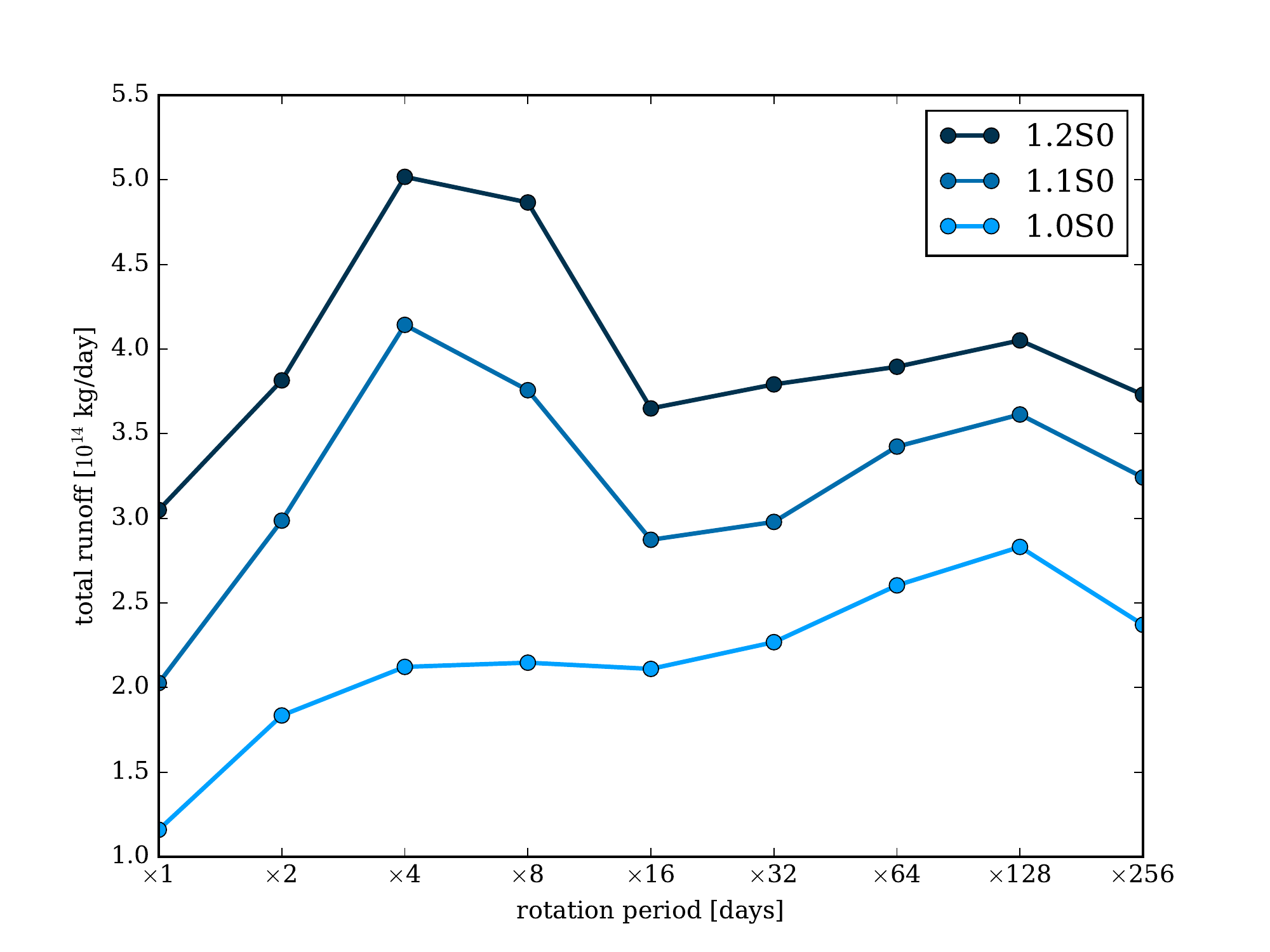}
    \caption{Total runoff in kg day$^{-1}$ as a function of rotation period and insolation S0.}
    \label{fig:runoffs}
    \end{minipage}
\end{figure*}

Because neither the average soil temperatures nor total runoffs display the same relation to rotation period as does the relative weathering rate (Figures \ref{fig:soil_temps} and \ref{fig:runoffs}), it is apparent that weathering rates cannot be solely attributed to a global average soil temperature nor amount of precipitation/runoff, but rather to a spatial conjunction of the two. For models with very long days, a soil depth greater than the current depth of 3.5 meters may be needed to sufficiently capture diurnal soil temperatures.

%% file: temp_hab_table.tex
\setlength{\tabcolsep}{10pt}

\begin{table}[t!]
  \centering
  \begin{tabular}{|p{2.8cm}|c|c|c|}
    \hline
    \multirow{2}{3cm}{\textbf{Rotation period [days]}} & \multicolumn{3}{c|}{\textbf{Insolation [S0]}} \\
    \cline{2-4}
    & 1.0 & 1.1 & 1.2 \\
    \hline
    1 & 0.76 & 0.97 & 0.98 \\ \hline
    2 & 0.78 & 0.97  & 0.98 \\ \hline
    4 & 0.80 & 0.97 & 0.98 \\ \hline
    8 & 0.81 & 0.92 & 0.98  \\ \hline
    16 & 0.92 & 0.98 & 0.98  \\ \hline
    32 & 0.87 & 0.96 & 0.99  \\ \hline
    64 & 0.78 & 0.88 & 0.94  \\ \hline
    128 & 0.74 & 0.82 & 0.87  \\ \hline
    256 & 0.71 & 0.75 &  0.82 \\ \hline
  \end{tabular}
  \caption{\label{table:hab_fracs}Fraction of the model's surface area where the surface temperature $T[^{\circ}$C] falls into the range $0\leq T\leq 100$, i.e.  `fractional habitability'. For reference, the Earth had a net fractional habitability of 0.85 according to temperatures measured in 2004 \citep{spiegel:2008}, although we reiterate that our models adopt a 0$^{\circ}$ obliquity and an eccentricity of 0, which should therefore not be directly compared to Earth values.}
\end{table}

%% file: discussion.tex
For Earth-like planets with rotation periods of 1, 2, 4, 8, 16, 32, 64, 128, and 256 days, we find for a 10\% and 20\% increase in insolation that average global surface temperatures decrease for planets with longer days (as shown in \citealt{way:2018b}). This drop in mean global surface temperature may be due to an increase in albedo from an increase in cloud cover at the substellar point, as has been observed in \cite{way:2018b} and in previous GCM studies of slowly rotating planets (e.g. \citealt{yang:2014}). In the near future we will continue to run these ROCKE-3D simulations to obtain sub-day averages in order to track cloud distribution and evolution with changing rotation periods.

For all insolations we find that surface temperatures become more uniform in latitude with increasing day lengths. This is due in part to a weakening Coriolis force with longer rotation periods, which enables the Hadley cells to grow in latitudinal extent, transporting heat more efficiently from the equator to the poles. This result is in qualitative agreement with previous studies that used GCMs to observe the effect of rotation rate on terrestrial planet atmospheres (e.g. \citealt{williams:1982, delgenio:1987, kaspi:2015, showman:2013} and references therein). Additionally, \cite{way:2018b} show by comparing ROCKE-3D simulations with a dynamic ocean (as used in this study) to ROCKE-3D simulations with a static ocean that the dynamic ocean significantly aids poleward heat transport and is responsible for a diminished ocean-ice fraction compared to the static ocean model.

Our results suggest that there may be a `rotational Goldilocks zone' in which the fractional habitability reaches a maximum at a critical rotation period (\S2.2, Figure \ref{fig:hab_frac}). We speculate that this critical rotation period occurs at the transition between two rotational regimes: one in which the efficiency of poleward heat transport increases with slower rotations, effectively increasing the surface temperatures in the upper latitudes; the other regime encompasses even slower rotations, where the Hadley cell has extended to the poles and the nightside significantly cools with increasingly long nights, while cloud coverage (and therefore albedo) continues to grow about the substellar point, effectively decreasing the mean global surface temperature.

Our simulation results for the solar insolation case show a peak in fractional habitability at a rotation period of 16 days (Figure \ref{fig:hab_frac}). Although there is no similar peak in the models with higher insolations, they do show a similar monotonic drop in fractional habitability with rotation periods beyond $\sim 16-32$ days. If these results are supported with additional models with finer rotation and insolation sampling, this could suggest a potential `Goldilocks' zone for planetary rotation rate in addition to orbital range. Given the expectation that rotation rate evolves due to tides, this also suggests the possibility of peak fractional habitability at a predictable system age. 

Across all insolations we also find that there is a peak in the rate of silicate weathering relative to the 001X1.0S0 model centered around a 4-day rotation period (Figure \ref{fig:weathering}). While the simulations used here maintain a fixed atmospheric CO$_{2}$ abundance, we can nonetheless draw some conclusions about the potential evolution of weathering and climate states. The carbon-silicate (C-S) cycle on the modern Earth operates with a weathering or sequestration rate of $\sim 7\times 10^{12}{\rm\ mols \;C\;yr^{-1}}$ (moles of carbon per year, e.g. \citealt{rushby:2018, brantley:1995}) and an estimated current mantle outgassing rate of $\sim 3\times 10^{12} {\rm\ mols \;C\;yr^{-1}}$ (e.g. \citealt{sleep:2001, zhang:1993}). Overall, together with other cycling pathways such as tectonic erosion and arc volcanism, it is estimated that the in-balance carbon fluxes in the Earth system operate at a level around $10^{12}-10^{13} {\rm\ mols\;yr^{-1}}$. Given a present-day estimated atmospheric carbon content of $\sim 1.8\times 10^{14} {\rm\ mols}$ this implies that $\sim 1-10\%$ of the Earth's atmospheric carbon is undergoing cycling per year as part of the C-S balancing feedback loop. Therefore there is considerable sensitivity on short timescales to any forcings that affect elements of the cycle.
 
 Our results (Figure \ref{fig:weathering}) suggest that weathering rates can peak by factors of $\sim 8$ (for higher insolations) due to precipitation and temperature changes from rotational evolution. We stress that we do not track the full evolution of a single planet, i.e. we do not incorporate a C-S cycle in our models, but rather hold atmospheric CO$_2$ constant, and we do not evolve rotations within a model run. In a real system we might, for example, expect a co-evolution of rotation-altered weathering with atmospheric CO$_2$ that leads to a progressive lowering of CO$_2$ and global temperatures that nonetheless remains in equilibrium at any given time due to the feedback cycle. However, this depends critically on the rate of rotational evolution. The overall climate state should adjust to daylength changes on comparatively rapid timescales (e.g. centuries), whereas the C-S cycle is completed over $\sim 10^8$ yr timescales \citep{sleep:2001}. 
 
 There is a wide range of expected rotational evolution timescales for worlds in the nominal habitable zone of different stellar masses, ranging from very rapid tidal evolution toward synchronous or pseudo-synchronous rotation (e.g. $<10^7$ years for low-mass stars \citealt{goldreich:1966, kasting:1993}) to timescales commensurate with a system's present age (e.g. $~10^9$ years as for a moonless Earth \citep{barnes:2016}). For terrestrial-analog planets at the inner edge of the liquid water orbital range, with higher than solar insolation, it is therefore possible that weathering can evolve to draw down atmospheric CO$_2$ at rates far in excess of C-S equilibrium fluxes.   

This could lead to climate plunging into a snowball state \citep{deitrick:2018} at intermediate rotation rates to those of the modern Earth and synchronous or psuedosynchronous rotators. In extreme cases a planet could remain snowballed until reaching its final slow rotation state. At this point whether or not it will `recover' to a more temperate state will hinge on the geophysical gas flux and the still poorly understood properties of climate on very slowly rotating worlds \citep{kite:2011}.

In light of our results, it is intriguing to speculate on the role of rotation in the faint young Sun paradox \citep{sagan:1972}, given the expectation for shorter day lengths on the early Earth. An exploration of this speculation however would require a new suite of simulations for the early Earth, with the appropriate solar insolation, range of atmospheric abundances, and shorter rotation periods.

 It is important to note that the exact relationship between the weathering rate and rotation is specific to a planet's landmass distribution. As we have shown, an Earth-like planet with a similar continental distribution to the Earth will experience an increased weathering rate for rotation periods $<$16 days as the Hadley cell, and therefore location of rainfall, expands to higher Northern latitudes where there is more landmass to be weathered. But for a planet with a different continental distribution, the rotation period which maximizes the silicate weathering rate could very well be shifted. Nonetheless, our results support the suggestion of \cite{rushby:2018} to consider the relationships between the carbonate-silicate cycle and fundamental planetary properties and the subsequent impact on long-term habitability.

%% file: conclusion.tex
In this study we have used the Goddard Institute for Space Studies' ROCKE-3D general circulation model to examine how increases in insolation and planetary rotation period may affect a planet's `fractional habitability' and silicate weathering rate. In light of our results, we argue that the rotation period and length of a planet's day is an important factor to consider in the determination of the habitability of exoplanets.

We believe that our present GCM study is the first to investigate the role of planetary rotation period on the silicate weathering rate. We have shown that the irradiation and length of an Earth-like planet's day has a significant effect on its draw-down rate of atmospheric CO$_{2}$. 

For both fractional habitability and silicate weathering we find evidence for peak-like behavior as a function of planetary day length. In terms of fractional habitability we suggest that there may be a `Goldilocks' zone for planetary rotation rate in addition to orbital location. We also propose that a young terrestrial-type planet experiencing fast rotational evolution due to star-planet tides may pass through an epoch of significantly enhanced CO$_{2}$ draw-down, potentially destabilizing its climate.

While it is extremely challenging to obtain observational constraints on rocky-exoplanet rotation rates, methods such as phase-curve photometry (e.g. \citealt{ford:2001}) might eventually yield results. We also suggest that a distinct climate state (such as low-temperature snowball states for water-rich worlds), combined with estimates of system age and tidal evolution, could itself conceivably provide clues to planetary spin.